\documentclass[aps,twocolumn]{revtex4}
\usepackage{epsfig}
\usepackage{amssymb}
\usepackage{amsmath}
\usepackage{amsfonts}
\newcommand{\be}{\begin{equation}}
\newcommand{\ee}{\end{equation}}
\newcommand{\ba}{\begin{eqnarray}}
\newcommand{\ea}{\end{eqnarray}}

\def\stau{{\widetilde \tau}}

\begin{document}

\title{Cosmogenic Neutrinos and Quasi-stable
Supersymmetric Particle Production}
\author{M. H. Reno}
\affiliation{%
Department of Physics and Astronomy, University of Iowa,
Iowa City, Iowa 52242 USA}
\author{I. Sarcevic and J. Uscinski}
\affiliation{%
Department of Physics, University of Arizona, Tucson, Arizona 85721,
USA}
\begin{abstract}

We study the signal for the detection 
 of quasi-stable supersymmetric particle produced 
in interactions of cosmogenic neutrinos.  We consider energy loss 
of high energy staus due to photonuclear and weak interactions.  We show that 
there are optimal nadir angles for which the 
stau signal is a factor of several hundred 
larger than muons.  We discuss how one could potentially eliminate 
muon background by considering the energy loss of muons in the detector.  
We also show results for the showers produced by weak interactions of 
staus that reach 
the detector.  

\end{abstract}

\maketitle

\section{INTRODUCTION}
%INTRODUCTION

Ultrahigh energy cosmic neutrinos could potentially probe 
physics beyond the Standard Model \cite{ringwald}.  Interactions of 
UHE neutrinos ($E_\nu \geq 10^{17}$eV) with nucleons
probe center of mass energies
above 14 TeV.  Some fraction of these neutrinos may 
produce 
supersymmetric particles or some other exotic particles.  
These processes are 
suppressed relative to standard model processes,
however, in some models interesting signals may arise from 
supersymmetric particles with long lifetimes.  In most 
SUSY scenarios, 
particles produced in high energy collisions 
decay immediately into the lightest one and are thus hard to 
detect.  
However in some low scale supersymmetric models in which 
gravitino is the lightest supersymmetric particle (LSP) and R-parity is 
conserved, the 
next-to-lightest particle (NLSP) is the charged superpartner of 
the right-handed tau, the stau 
 \cite{susy_models}. 
Due to its weak coupling to the gravitino, the stau is a long-lived 
particle in these models.  
For the 
supersymmetric breaking scale, 
$\sqrt F >5\times10^6$ GeV, 
the long-lived stau 
could travel distances of the order of $10^4$ km 
before decaying into the 
gravitino.  

The distance that staus travel before decaying depends on the 
gravitino mass (or 
equivalently on the supersymmetry breaking scale) and the stau mass.  
Limits on the stau mass of about 100 GeV 
come from its non-observation in  
accelerator experiments 
\cite{stmass1,stmass2,stmass3,stmass4}.  
Recently it was proposed that staus produced in high energy neutrino 
interactions, where neutrinos originate in astrophysical 
sources, might be detectable in 
neutrino telescopes \cite{ABC,Ahlers}.  

The cross section for the production of staus in neutrino-nucleon scattering 
\cite{ABC} is 
several orders of magnitude smaller than the neutrino charged-current or 
neutral-current cross section \cite{gqrs}.   However, 
once produced, the long-lived 
staus have the potential to travel through the earth  without decaying 
and thus open up a possibility to be detected in 
neutrino telescopes.  The long range of staus could potentially  
compensate for the suppression in the production cross section 
by increasing the effective detector volume and therefore enhancing the
signal.

The detection of staus depends on the stau lifetime and range, so
it is important to determine the  
energy loss as it traverses the earth.  
The details of the range depend in part
on the supersymmetry breaking and how the quasi-stable stau particle
is comprised of the SUSY partners of the right-handed and left-handed
taus. The electromagnetic energy loss has been shown to 
have the largest contribution from photonuclear interactions for stau energies
between $10^6$-$10^{12}$ GeV, resulting in a range of $10^4$ km.w.e. for masses
of the order of a few hundred GeV \cite{RSS}.  
Weak interactions may come into play as well.
The stau range has been shown to be 
sensitive to 
the mixing angle of right-handed and left-handed staus.  When the mixing is
maximal, weak interactions act to suppress the range at energies above 
$\sim10^9$ GeV \cite{RSU1}, however, their weak interactions have the potential 
to produce signals in neutrino 
detectors
such as the Antarctic Impulse Transient Array (ANITA) \cite{anita} and 
the Antarctic Ross Iceshelf Antenna Neutrino Array (ARIANNA) \cite{arianna}.

The high energies required for stau production lead us to 
focus on the production of staus in interactions of 
cosmogenic neutrinos as 
they traverse the Earth and/or in the detector.  
 These neutrinos originate from cosmic ray protons 
interacting with the cosmic microwave background,
$$ p\gamma(3{\rm K}) \rightarrow \Delta \rightarrow
N\pi$$
followed by charged pion, muon and neutron decays.  
This flux is guaranteed as cosmic ray fluxes are measured as well as the
3K microwave background.  We use a conservative cosmogenic neutrino
flux evaluated by Engel, Seckel and Stanev (ESS) in Ref. \cite{ess}.
They evaluate the neutrino flux associated with the 
 measured cosmic ray flux by tracing back 
 cosmic ray propagation through the background radiation.  
Depending on the cosmological evolution assumed, the overall 
normalization of this flux has an uncertainty of 
about a factor of four.   
In addition, neutrinos 
could be produced at the sources of the high energy cosmic
rays and those are not included in the evaluation of ESS neutrino flux.  
 Thus, the ESS neutrino flux is 
a conservative estimate of the cosmogenic flux.  
  
The cosmogenic neutrino 
flux, when neutrino flavor oscillations are not 
included, peaks at high energies, around $10^8$ GeV, 
 and thus it is in the energy range where ANITA and 
ARIANNA have very good sensitivity \cite{anita,arianna}.  
The neutrino flavor ratio for cosmogenic neutrinos deviates from the 
common 2:1 ratio, due to the neutron decay contribution to electron neutrinos 
\cite{ess}.  
This implies that one needs to consider three flavor oscillations when 
considering the cosomogenic neutrino flux that arrives at the Earth \cite{us_JMRS}.  

We consider cosmogenic neutrinos, their propagation, stau production,
and subsequent energy loss as it traverses the earth, for a region of 
parameter space where the staus do not decay over the distances required.  
We compare the
resulting stau flux when there is no mixing between right-handed and 
left-handed stau and when there is maximal mixing.  
We consider muon-like signals (charged tracks) produced by staus and its 
associated background.  
We discuss the potential for eliminating the background by 
measuring the energy loss, which requires large volume detectors.   
Finally we discuss the 
showers produced in the ice due to stau interactions and its background 
from neutrino-induced showers.  

\section{Characteristic Distance Scales and Stau Production}

\begin{figure}[t]
\begin{center}
\epsfig{file=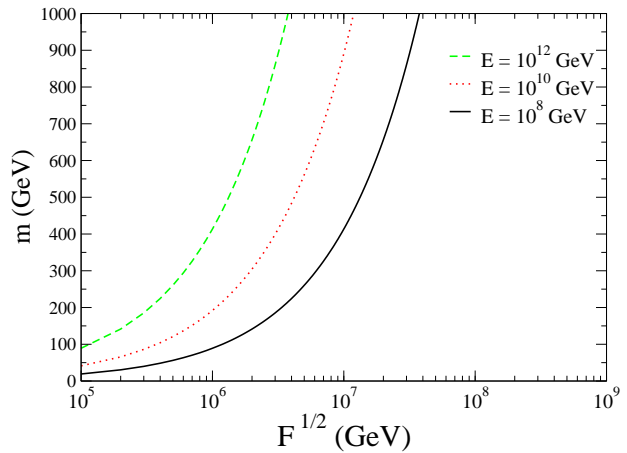, width=2.75in,angle=270}
\end{center}
\caption{The parameter space of stau mass $m$ and decay parameter
$F^{1/2}$  probed with stau decay lengths corresponding to 
a column depth of $10^7$ cm w.e.  The area to the 
the right of each curve is the parameter space probed by 
demanding that stau does not 
decay as it traverses this column depth.}
\label{fig:paramspace}
\end{figure}

The potential for staus to have a long lifetime is an essential
feature of this analysis. The
lifetime of the stau depends on the supersymmetry breaking scale 
($F^{1/2}$)
and stau mass ($m_\stau$).
It is given by
\begin{eqnarray}
c\tau=\Bigl(\frac{F}{10^{14}\ {\rm GeV}^2}\Bigr)^2 \Bigl(\frac{100 {\rm GeV}}{m_\stau}
\Bigr)^5 10\ {\rm km}\, .
\end{eqnarray}
In Fig. \ref{fig:paramspace}, 
we show the parameter space ($m_\stau, \sqrt F$) and the region for
which staus with certain energy ($10^8,\ 10^{10}, \ 10^{12}$ GeV)
have long enough lifetimes, 
$\gamma c\tau = 10^7$ cm w.e., 
to travel through the Earth without
decaying for a nadir angle of $88^\circ$.  
Different nadir angles, or column depths, correspond to
 different
range of these parameters.  Clearly there is a large parameter space in 
 which staus do not decay, but they may interact via charged-
current interactions, if there is a maximal mixing between the right-handed and
left-handed stau. 
Here we consider
the case when stau mass is 150 GeV and the supersymmetric breaking scale F$^{1/2}$
=10$^7$ GeV.

\begin{figure}[t]
\begin{center}
\epsfig{file=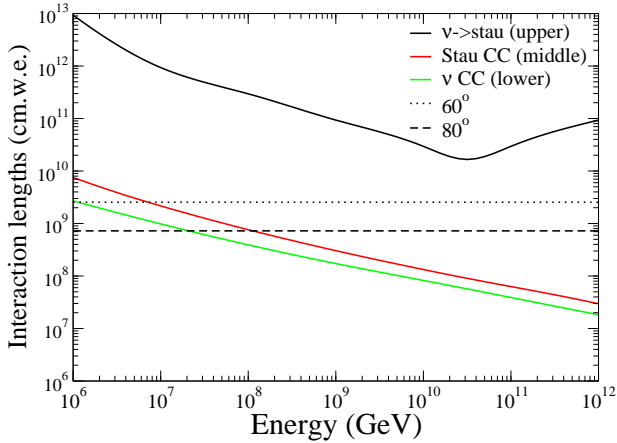,width=2.75in,angle=270}
\end{center}
\caption{Interaction lengths for neutrinos and staus. The solid
curves, from top to bottom, are the interaction lengths for neutrinos
to produce staus, for stau charged current (CC) interactions
with maximal weak interactions and
for neutrino CC interactions. The dotted and dashed lines show the column
depths for nadir angles of $60^\circ$ and $80^\circ$.}
\label{fig:intlength}
\end{figure}

Staus are produced in neutrino-nucleon interactions in the Earth with
the cross sections\cite{ABC} about three orders of magnitude
smaller than standard model
neutrino weak interactions with nucleons \cite{gqrs}.
The staus are produced by the decays of heavy quarks and sleptons
produced in the primary neutrino-nucleon interactions. 
The cross section for the stau production is a t-channel production of
a slepton and a squark via gaugino exchange.  This cross section
depends on the squark mass,
gaugino masses and the mass of the left-handed slepton. 
We take the cross section
for stau production obtained with $m_{\widetilde q}=300$ GeV, 
$m_{\widetilde w}=250$ GeV and
$m_{\widetilde l_L}=250$ GeV \cite{ABC}. 
Stau cross sections obtained using different parameters and different SUSY 
models are shown in Ref.\cite{Ahlers}.  
% We discuss later how different choice of parameters affects our results.
The cross section converted into an interaction length,
\begin{equation}
{\cal L} = (N_A\sigma)^{-1}\ ,
\end{equation} 
is shown for neutrino production of staus by the upper solid line
in Fig. \ref{fig:intlength}.

Cosmic neutrino fluxes that reach the Earth get attenuated as neutrinos 
traverse the Earth toward the detector, due to weak interactions, 
primarily the 
charged-current (CC) interactions.  
Similarly, once created in neutrino-nucleon interactions, the 
stau flux also 
gets attenuated when staus interact weakly.   The size of these effects 
can be seen in 
Fig. \ref{fig:intlength}, where we show neutrino interaction length \cite{gqrs}
and the 
stau interaction length \cite{RSU1} due to their
charged-current interactions with isoscalar nucleons, as a
function of incident energy.  This figure shows 
 the stau charged current cross section
with maximal weak interactions.  We also show for reference 
the column depth
for nadir angles of $60^\circ$ and $80^\circ$.

\section{Lepton and Stau Fluxes}

\subsection{Cosmogenic Flux}

The cosmogenic fluxes that we use as reference input fluxes are the
standard evolution and strong evolution fluxes of Engel, Seckel and
Stanev \cite{ess}.  The fluxes are of neutrinos and antineutrinos,
with an electron neutrino to muon neutrino flavor ratio that depends
on energy. Part of this energy dependence comes from the eventual
neutron decay in $p\gamma \rightarrow n\pi^+$ production.

Neutrino oscillations due to non-zero neutrino masses modify the neutrino
flavor ratio over the cosmological distances they travel. Consequently,
we use for the cosmogenic fluxes at the Earth \cite{us_JMRS},
\begin{eqnarray}
F_{\nu_e} &=& F_{\nu_e}^0-\frac{1}{4}\sin^2 2\theta_{12} \Bigl( 2 F_{\nu_e}^0
- F_{\nu_\mu}^0 - F_{\nu_\tau}^0 \Bigr) \\
F_{\nu_\mu} &=& F_{\nu_\tau} = \frac{1}{2}\Bigl( F_{\nu_\mu}^0 + F_{\nu_\tau}^0 \Bigr)\nonumber \\
&& + \frac{1}{8}\sin^2 2\theta_{12} \Bigl( 2 F_{\nu_e}^0
- F_{\nu_\mu}^0 - F_{\nu_\tau}^0 \Bigr),
\end{eqnarray}
where $F_i^0$ are the ESS cosmogenic fluxes \cite{ess}.
 The cosmogenic
tau neutrino flux $F_{\nu_\tau}^0=0$ because of the threshold
for tau production: the dominant process of resonant $\Delta$ production
ultimately results in only $\nu_e$ and $\nu_\mu$ fluxes at the
point of production.
The expressions in Eqs. (3) and (4) are written with the assumption of
maximal mixing between $\nu_\mu$ and $\nu_\tau$. For the numerical work
below, we have taken $\theta_{12}=33^\circ$.

\begin{figure}[t]
\begin{center}
\epsfig{file=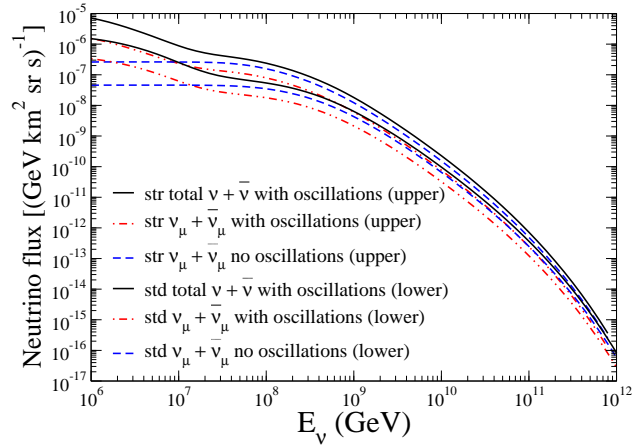,width=2.75in,angle=270}
\end{center}
\caption{Cosmogenic neutrino flux with
standard evolution and with strong evolution \cite{ess}.  
The upper (lower) solid lines represent the 
total neutrino plus antineutrino flux 
for strong (standard) evolution, dashed lines 
are the $\nu_\mu+\bar{\nu}_\mu$
flux without including oscillation effects, and the dot-dashed lines represent 
$\nu_\mu+\bar{\nu}_\mu$ flux with oscillation effects included.} 
\label{fig:nulfluxin}
\end{figure}

In Fig. \ref{fig:nulfluxin}, the total neutrino
plus antineutrino cosmogenic fluxes are shown
with the upper (strong evolution) and lower (standard evolution)
solid lines. Without neutrino oscillations, the corresponding
$\nu_\mu+\bar{\nu}_\mu$
fluxes are shown by the dashed lines. Including neutrino
oscillations, the $\nu_\mu+\bar{\nu}_\mu$ cosmogenic strong
and standard evolution fluxes are shown by the dot-dashed  
lines. Neutrino oscillations enhance the incident
muon (anti-)
neutrino flux at low energies due to $\bar{\nu}_e\rightarrow
\bar{\nu}_\mu$
oscillations.

Finally, as noted above, 
the cosmogenic neutrino fluxes shown in Fig. \ref{fig:nulfluxin}
are considered ``conservative.''   They correspond to 
neutrinos produced by cosmic rays interacting with the 3K
radiation.  Neutrinos could also be produced
in some cosmic accelerators, such as active galactic nuclei and 
gamma ray bursts, but the fluxes have large theoretical 
uncertainties \cite{uhenufluxes}.  We focus on the cosmogenic neutrinos
from cosmic ray interactions with the microwave background because they 
are a ``guaranteed'' source of ultrahigh energy neutrinos.  

%%%%%%%%%%%%%%%%%%%%%%%%%%%%Muon Flux%%%%%%%%%%%%%%%%%%%%%%%%%%%%%%%%%
\subsection{Muon Flux}

Muons are produced by $\nu_\mu N$ charged-current interactions.
Electromagnetic energy loss of muons passing through matter
together with a survival probability that depends on lifetime
and energy determines the muon flux. Electromagnetic
energy loss is described by the average change in energy per unit
column depth $X= \rho z$ (in terms of density $\rho$ and
distance $z$) via
\begin{equation}
\label{eq:dedx}
\frac{dE}{dX} = -(\alpha +\beta_\mu E)\ .
\end{equation}
The radiative energy loss due to bremsstrahlung, pair production and 
photonuclear scattering, characterized by 
$\beta_\mu$, increases 
with energy from about 
$ \simeq 4\times 10^{-6}$ cm$^2$/g for $E_\mu \sim 10^3$ GeV to about 
$ \simeq 5-6\times 10^{-6}$ cm$^2$/g for $E_\mu \sim 10^9$ GeV \cite{drss}.  
Here we take 
$ \beta_\mu \simeq 4\times 10^{-6}$ cm$^2$/g, which will correspond to 
the maximum background for the stau charged-track signal due to muons.  
 With $\alpha\simeq 2\times 10^{-3}$ GeV\,cm$^{2}$/g \cite{ionization},
for the energies considered here, $dE/dX\simeq - \beta_\mu E$.
Eq. (\ref{eq:dedx}) combined with effects of the finite muon
lifetime on the survival probability $P_{\rm surv}$,
\begin{equation}
\label{eq:dpdx}
\frac{dP_{\rm surv}}{dX} = -\frac{P_{\rm surv}}{c\tau\rho E/m}
\end{equation}
leads to
\begin{equation}
P_{\rm surv} = \exp
\Biggl[-\frac{m_\mu}{c\tau_\mu \beta_\mu\rho}\Biggl( \frac{1}{E_\mu}
-\frac{e^{-\beta_\mu (L-X)}}{E_\mu}\Biggr)\Biggr]\ .
\end{equation}

The muon flux produced by incident muon neutrinos, for column depth $L$ is given 
by 
\begin{eqnarray}
\nonumber
F_\mu(E_\mu,L)  &\simeq &  \int_0^L dX \int dE_\nu e^{ {-X\sigma_{CC}(E_\nu) N_A}} F_{\nu_\mu}(E_\nu,0)\\ \nonumber
& \times & N_A {\sigma_{\nu\rightarrow \mu}(E_\nu)}
\delta(E_\mu -{0.8} E_\nu e^
{-\beta_\mu (L-X)})\\
&\times & \exp
\Biggl[-\frac{m_\mu}{c\tau_\mu\beta_\mu\rho }\Biggl( \frac{1}{E_\mu}
-\frac{e^{-\beta_\mu (L-X)}}{E_\mu}\Biggr)\Biggr]\ ,
\label{eq:fmu}
\end{eqnarray}
where we have made some simplifying approximations.
We have approximated the neutrino-nucleon charged current differential
cross section by
$$\frac{d\sigma _{\nu\rightarrow \mu}(E_\nu,\ E_\mu')}{dE_\mu'}\simeq
\sigma_{\nu\rightarrow \mu}(E_\nu)\delta (E_\mu'-0.8 E_\nu)$$
where $E_\mu'$ is the energy of the produced muon. This
is the energy of the muon before it
loses energy via electromagnetic interactions.
The flux $F_\nu$ represents the $\nu_\mu + \bar{\nu}_\mu$ flux,
and $F_\mu$ is the sum of muon and antimuon fluxes.

We have also approximated the attenuation
of the neutrino flux in transit through a column depth $X$ by
the shadow factor
\begin{equation}
S\equiv \exp \Bigl( -\sigma_{CC}^{\nu N}(E_\nu) N_A X \Bigr)\ .
\end{equation}

The muon fluxes, as a function of muon energy, are shown in Figs.
\ref{fig:fluxosc-std} and \ref{fig:fluxosc-str} with dotted lines
for two nadir angles, $80^\circ$ and $88^\circ$. Also shown on these
figures are the stau fluxes, to which we now turn.

\subsection{Stau Fluxes}

The stau flux depends on a number of inputs. We use the squark
and slepton masses indicated in Sec. II yielding the 
$\nu\rightarrow \stau$ interaction
length shown in Fig. 2. A further choice of parameter is the
mixing angle between weak isospin zero and weak isospin 1/2 scalars,
the right-handed and left-handed staus, that yields the mass
eigenstate of the NLSP stau. 
We consider two limiting cases: staus with no weak interactions,
and staus with maximal weak interactions \cite{RSU1}. 

We start with the case of no
weak interactions.
In this case, the
evaluation of the stau flux is similar to that of the muon flux.
In Ref. \cite{RSS}, we showed that
for staus, a reasonable parametrization of the
energy loss parameter $\beta_\stau$ is
\begin{eqnarray}
\label{eq:staubeta}
\beta_\stau &=& b_0+b_1\ln
(E/E_0)\ ,\\ 
\nonumber b_0 &=& 5\times 10^{-9}\ {\rm cm}^2/{\rm g} \ ,
\\ \nonumber 
b_1 &=& 2.8\times 10^{-10}\ {\rm cm}^2/{\rm g} \ ,
\\ \nonumber 
E_0 &=& 10^{10}\ {\rm GeV}\ .
\end{eqnarray}
With the expression in Eq. (\ref{eq:staubeta}), the stau
survival probability and the relation between the initial
stau energy and the final stau energy as a function of distance
is modified from the constant $\beta$ case, namely \cite{Dutta:2005yt} 

\begin{eqnarray}
\nonumber
E_{\widetilde \tau}^i(E_{\widetilde \tau}) &=& E_0 \exp\Biggl[\Bigl[
 \frac{b_0}{b_1}(1-e^{-b_1 (L-X)})\\
&+&\ln\frac
{E_{\widetilde \tau}}{E_0}
\Bigr] e^{b_1(L-X)}\Biggr]
\label{eq:etili} \\
\nonumber
P_{\rm surv}(E_\stau ,E_\stau^i ) &=& \exp\Biggl(\frac{m_\stau b_1}{
c\tau\rho b_0^2}\Biggl[\frac{1}{E_\stau}(1+\ln(E_\stau/E_0) \\ \nonumber
&-& \frac{1}{E_\stau^i}(1+\ln(E_\stau^i/E_0)\Biggr]
\Biggr)\\
\label{eq:psurvstau}
&\times & \exp\Biggl[ -\frac{m_\stau}{c\tau b_0\rho}
\Biggl(\frac{1}{E_\stau} -\frac{1}{E_\stau^i}\Biggr)\Biggr]
\end{eqnarray}
This leads to a stau flux of
\begin{eqnarray}
\nonumber
F_\stau(E_\stau,L)  &\simeq &  2\int_0^L dX \int dE_\nu e^{ {-X\sigma_{CC}(E_\nu) N_A}} F_\nu(E_\nu,0)\\ \nonumber
& \times & N_A {\sigma_{\nu\rightarrow \stau}(E_\nu)}
\delta(E_\stau -\frac{1}{6} E_\nu  \frac{E_\stau}{E_\stau^i(E_\stau)})\\
&\times & P_{\rm surv}(E_\stau,E_\stau^i)
\ .
\label{eq:ftaubc}
\end{eqnarray}
The prefactor of 2 accounts for
the fact that the staus appear in pairs, one from each of the chain
of decays of the initial squark and slepton. All of the neutrino
plus antineutrino flavors are included in 
$F_\nu(E_\nu,0)$.

\begin{figure}[t]
\begin{center}
\epsfig{file=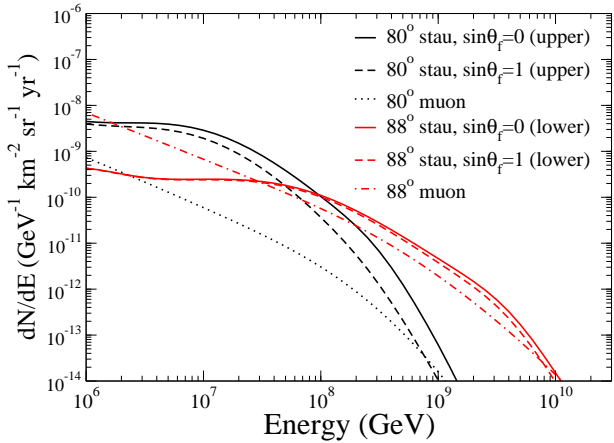,width=2.75in,angle=270}
\end{center}
\caption{For fixed nadir angles of $80^\circ$ and $88^\circ$,  
the stau flux assuming no weak interactions of the staus (solid lines)
and assuming maximal weak interactions (dashed lines)
produced by the ESS neutrino flux with standard evolution,
evaluated using Eq. (\ref{eq:ftaubc}) and an input flux with including
oscillations.
The dotted line shows the neutrino induced muon flux at $80^\circ$. 
}
\label{fig:fluxosc-std}
\end{figure}

\begin{figure}[t]
\begin{center}
\epsfig{file=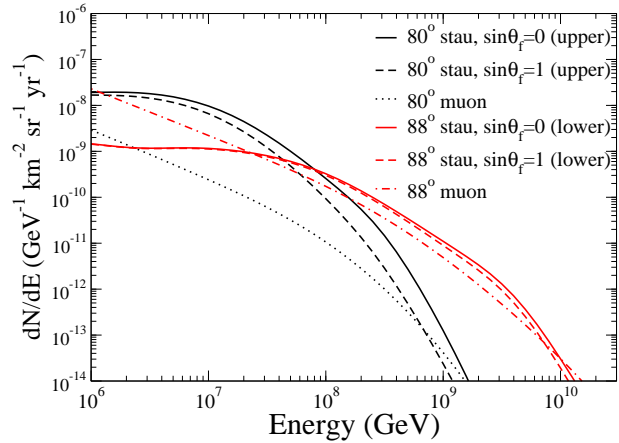,width=2.75in,angle=270}
\end{center}
\caption{For fixed nadir angles of $80^\circ$ and $88^\circ$, 
the stau flux assuming no weak interactions of the staus (solid lines)
and maximal weak interactions (dashed lines)
produced by the ESS neutrino flux with strong evolution,
evaluated using Eq. (\ref{eq:ftaubc}) and an input neutrino 
flux with oscillations.
The dotted line shows the neutrino induced muon flux at $80^\circ$. 
}
\label{fig:fluxosc-str}
\end{figure}

When weak interactions are included, there is an effect due to
the attenuation of the staus themselves. The survival probability
is modified and is given by 
\begin{eqnarray}
\label{eq:dpdxmod}
\frac{dP_{\rm surv}}{dX} &=& -\frac{P_{\rm surv}}{\lambda_{\rm eff}}\\
\lambda_{\rm eff}^{-1} &=& (c\tau\rho E/m_{\widetilde \tau})^{-1}
+ N_A \sigma^{CC}({\widetilde \tau} N)
\end{eqnarray}
The solution of the combined equations Eqs. (\ref{eq:dpdxmod}) and
(\ref{eq:dedx}) is done numerically.

The resulting stau fluxes, for fixed nadir angles of $80^\circ$, 
$85^\circ$ and
$88^\circ$ are shown as a function of stau energy in Fig.
\ref{fig:fluxosc-std} for the ESS standard evolution flux, and in
Fig. \ref{fig:fluxosc-str} for the strong evolution.
In these figures, the solid lines are with minimal weak interactions,
while the dashed lines are for maximal weak interactions.

\section{Stau signals}

\subsection{Stau Charged Tracks}

The fluxes of staus and muons depend on the initial neutrino fluxes, the
production cross sections, and the energy loss parameter $\beta$.  
As noted in Sec. III, for the
stau signal, the initial neutrino flux includes all 
neutrino flavors while for muons, only the oscillated 
$\nu_\mu+\bar{\nu}_\mu$ flux contributes.  
The neutrino energy required to produce a stau or a 
muon
of comparable initial energy along the trajectory through the earth is 
about 5 times 
higher for the stau because the mean energy of the resulting stau is
approximately 1/6 of the incident neutrino energy, in contrast to
the muon case where $E_\mu'\sim 0.8 E_\nu$.  
This means we are probing higher energy neutrinos for 
staus
than for muons to produce a quasi-stable particle
of a given energy. On the other hand,
the muons lose more energy in transit from the production point
to the detection point. 

The production cross section for staus is
approximately three orders of magnitude smaller than the muon production
cross section.   Once the staus are 
produced, they lose very little energy as they traverse the earth while the 
muon energy loss is of the order $10^2-10^3$ times greater.  The stau range 
can be as high as $10^4$ km.w.e. for vanishing charged-current interactions
or suppressed to about $10^3$ km.w.e. for maximal charged-current 
interactions, both higher than the muon range.  The neutrino attenuation
will also have a large effect on the signals.  This attenuation 
acts to deplete more
muons than staus since the muons must be created very near 
the detector to be seen, whereas the staus can be produced farther away.

\begin{figure}[t]
\begin{center}
\epsfig{file=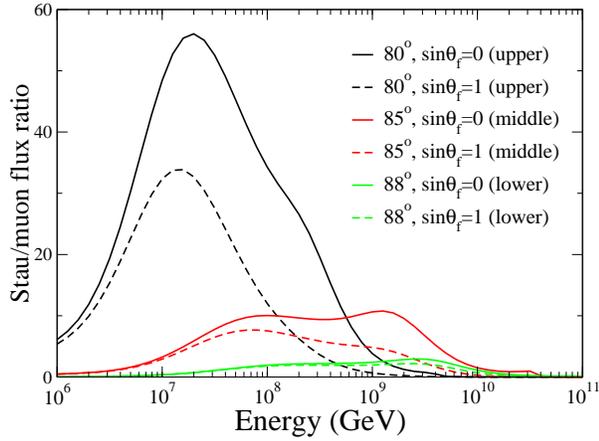,width=2.75in,angle=270}
\end{center}
\caption{For fixed nadir angles of $80^\circ$, $85^\circ$ and 
$88^\circ$,
the ratio of the stau flux assuming no weak interactions of the staus
(solid lines) and maximal weak interactions (dashed lines)
to the muon flux
produced by the ESS neutrino flux with standard evolution,
including oscillations.}
\label{fig:rat-std}
\end{figure}

These competing
effects account for the large stau/muon ratio for specific angles and
particle energies.  In Fig. \ref{fig:rat-std}, we show the ratio of the
stau flux to the muon flux 
for the angles $80^\circ$, $85^\circ$, and $88^\circ$, 
as a function of energy, for
standard ESS evolution.  
The solid lines show the ratio in the scenario where stau weak
interactions do not occur.
The effect of including maximal charged-current interactions
of the staus can be seen by the suppression in the ratio, shown with the
dashed lines.  We note that the suppression in the
ratio begins to take effect between $10^6$ and $10^7$ GeV, where
the stau range begins to be affected by stau charged-current interactions 
 \cite{RSU1}. 
In Fig. \ref{fig:rat-std}, the largest ratio occurs for the nadir angle of
$80^\circ$ which represents the largest path length through the earth of the
trajectories shown.  As the nadir angle is 
decreased, there is an enhancement in the ratio until it reaches a maximum
value and then it drops off when the neutrino attenuation begins to 
be the dominant effect for the signal. At lower energies, $E_\stau \sim 
2 \times 10^6$ GeV, the maximum ratio is 125 (at $70^\circ$) 
 when stau maximum weak interactions are 
included and it is 280 when there are no stau weak interactions.  
Furthermore, our signal to background ratios increase by about 
$50\%$ (for energy 
around $10^7$ GeV) if we 
incorporate energy dependent muon energy loss, by increasing 
$\beta_\mu$ value to $\beta_\mu \simeq 6\times 10^{-6}$ cm$^2$/g.  

\begin{figure}[t]
\begin{center}
\epsfig{file=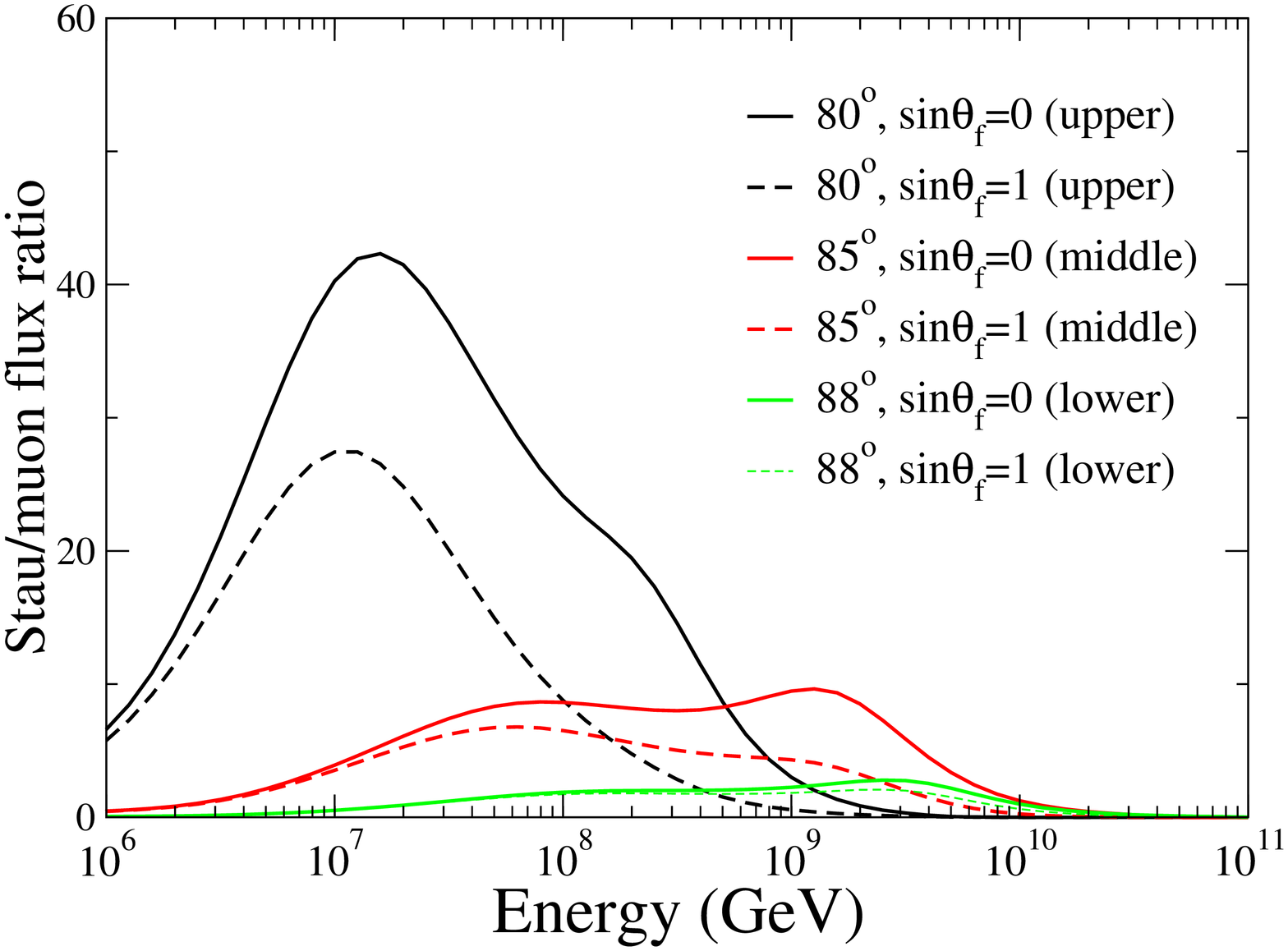,width=2.75in,angle=270}
\end{center}
\caption{For fixed fixed nadir angles of $80^\circ$, $85^\circ$ and 
$88^\circ$,
the ratio of the stau flux assuming no weak interactions of the staus (solid
lines) and maximal weak interactions (dashed lines)
to the muon flux produced by the ESS neutrino flux with strong evolution,
including neutrino oscillations.}
\label{fig:rat-str}
\end{figure}

\begin{figure}[t]
\begin{center}
\epsfig{file=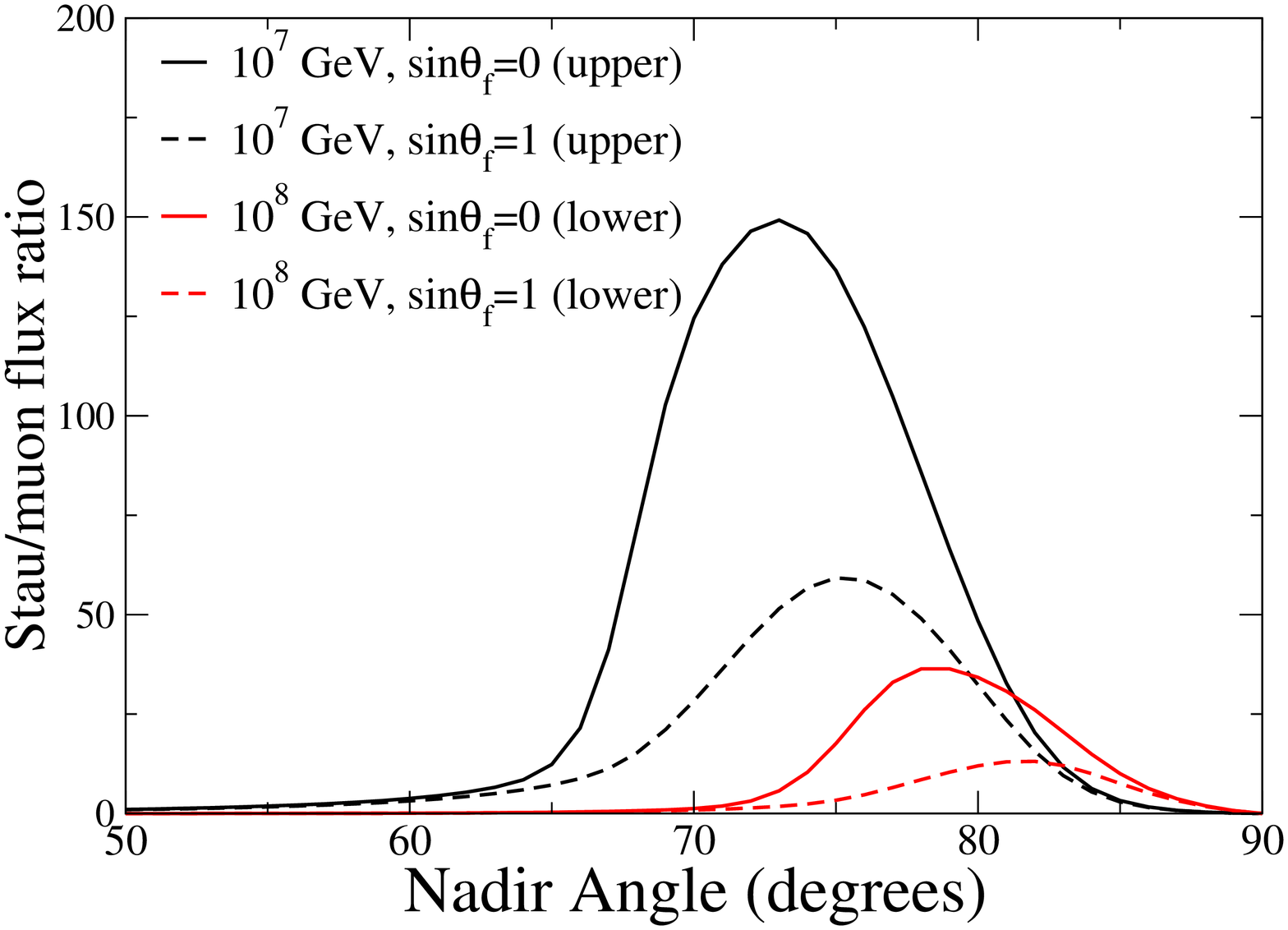,width=2.75in,angle=270}
\end{center}
\caption{For fixed energies of $10^7$ and $10^8$ GeV,
the ratio of the stau flux assuming no weak interactions of the staus
(solid lines) and maximal weak interactions (dashed lines)
to the muon flux
produced by the ESS neutrino flux with standard evolution,
including oscillations.}
\label{fig:ratoscvth-std}
\end{figure}

\begin{figure}[t]
\begin{center}
\epsfig{file=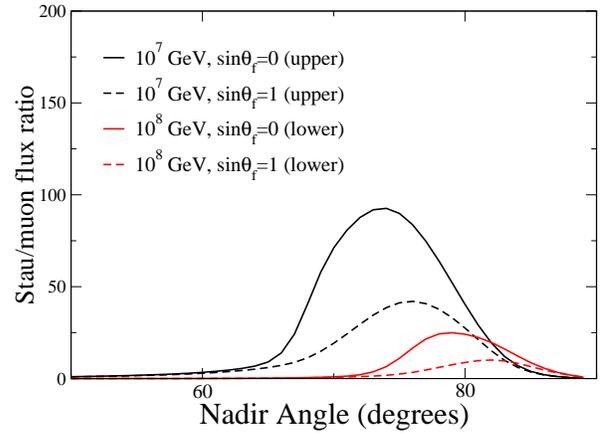,width=2.75in,angle=270}
\end{center}
\caption{For fixed energies of $10^7$ and $10^8$ GeV,
the ratio of the stau flux assuming no weak interactions of the staus (solid
lines) and maximal weak interactions (dashed lines)
to the muon flux produced by the ESS neutrino flux with strong evolution,
including neutrino oscillations.}
\label{fig:ratoscvth-str}
\end{figure}

Fig.
\ref{fig:rat-str} shows the same effects for strong ESS evolution
where the ratio is somewhat reduced.  We show in Figs.
\ref{fig:ratoscvth-std}  and \ref{fig:ratoscvth-str} the ratio for
stau/muon fluxes as a function of angle for the two fixed final particle energies
$10^7$ and $10^8$ GeV, for ESS standard and strong evolutions.  Again,
we see the same enhancement of stau/muon ratio shifted to 
the lower angles shown.  The strong ESS evolution shows the same behavior as 
standard evolution, again with a suppressed ratio.

Figures \ref{fig:rat-std}-\ref{fig:ratoscvth-str}
show a significant enhancement of the
stau flux relative to the muon flux for various energies and
angles. Extracting the stau signal, however, is quite difficult.
At issue is the fact that apart from stau decays and charged current
interactions, the stau moves through matter in a way similar to
a lower energy muon: the stau signal is muon-like.

In terms of electromagnetic energy loss, one of the issues is that
the average of the energy loss per unit distance at high energy
scales with $\beta E$. The electromagnetic energy loss parameter
$\beta$ for pair production and photonuclear contributions to the energy
loss scales as the inverse mass of the charged particle. The bremsstrahlung
process, important for muon energy loss, is negligible for
stau energy loss since it scales with the inverse mass squared.

\begin{figure}[t]
\begin{center}
\epsfig{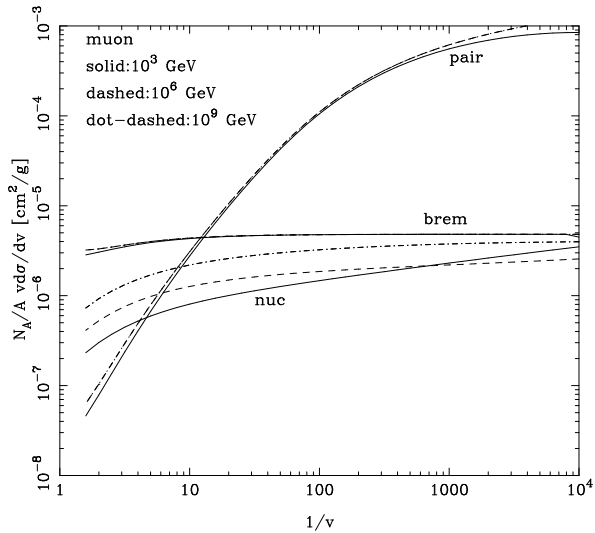}
\end{center}
\caption{The inelasticity weighted differential cross section
for muons for three muon energies: $10^3$ GeV (solid lines),
$10^6$ GeV (dashed lines) and $10^9$ GeV (dot-dashed lines) for
pair production, bremsstrahlung and photonuclear energy loss processes.
}
\label{fig:muon-vdsdv}
\end{figure}

An important element of the energy loss is the electron positron
pair production cross section. This cross section is important
at low values of the inelasticity parameter $v$ where
\begin{equation}
v=\frac{E-E'}{E}\ ,
\end{equation}
the ratio of the change in muon or stau energy to the initial energy.
In Fig. \ref{fig:muon-vdsdv} we show the differential cross section
$v d\sigma /dv$ scaled by $N_A/A$ for
muons scattering in iron \cite{koko,pairmeter}.
For ease of viewing the small $v$ region,
we plot the differential cross section versus $1/v$.
Because the photonuclear
process includes inelastic scattering corrections \cite{drss,bugaev},
the differential cross section depends on the incident muon
energy. Pair production dominates for $1/v$ greater than $\sim 10$, that
is to say for muon energy losses of less than $\sim 10\%$ of the
initial muon energy.

\begin{figure}[t]
\begin{center}
\epsfig{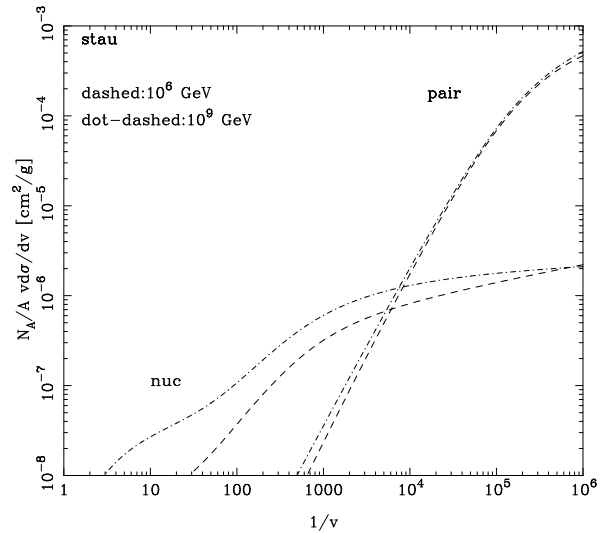}
\end{center}
\caption{The inelasticity weighted differential cross section
for staus for two stau energies:
$10^6$ GeV (dashed lines) and $10^9$ GeV (dot-dashed lines) for
pair production and photonuclear energy loss processes. Bremsstrahlung
energy losses are negligible for staus.
}
\label{fig:stau-vdsdv}
\end{figure}

The corresponding differential cross section for staus is shown in
Fig. \ref{fig:stau-vdsdv}. This figure shows that the dominant
process for energy loss of staus is pair production only for energy
losses of less than $v \sim 10^{-4}$. Similar results are obtained
for ice.

\begin{figure}[t]
\begin{center}
\epsfig{file=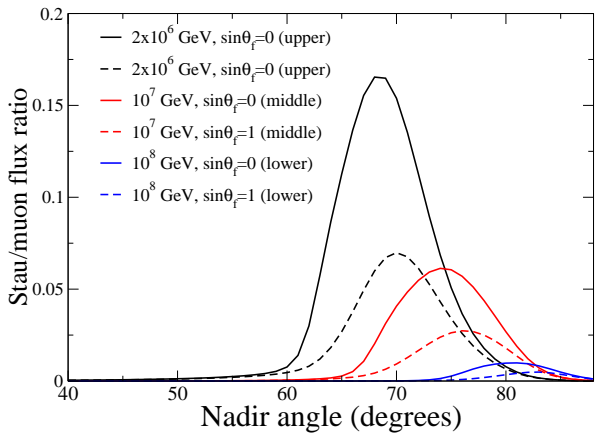,width=2.75in,angle=270}
\end{center}
\caption{ 
The ratio of stau flux at energy rescaled by $m_\stau/m_\mu$ as a function 
of the nadir angle for muon energy of 
$10^7$ GeV and $10^8$ GeV including stau maximal weak interactions (dashed 
lines) and no weak interactions (solid lines).}
\label{fig:ratstdeff}
\end{figure}

Figs. \ref{fig:muon-vdsdv} and \ref{fig:stau-vdsdv} can be
interpreted in terms of the average energy loss $\langle dE/dX\rangle$.
For a stau of energy $E_\stau$, the average energy loss per unit distance
is of order of that of a muon with energy $E_\mu\sim 10^{-3} E_\stau
\sim m_\mu/m_\stau\cdot E_\stau $. If the average energy loss per unit
distance is the only observable quantity, the large stau to
muon ratios are unobservable because one must compare the high
energy stau flux with the lower energy, but significantly larger,
muon flux. Quantitatively, this is shown in Fig. \ref{fig:ratstdeff}.
Here we take the ratio of the stau flux at energy $E_\stau =m_\stau/
m_\mu\cdot E_\mu$ to the muon flux evaluated
at $E_\mu$.  We note that the ratio is peaked at the same angle as 
in Fig. 8, but the ratio is much smaller.  
This is because $\langle dE/dX\rangle$ matches, for example, muons with
energies of $10^7$ GeV to staus
with energies more than  
two orders of magnitude higher, where the stau flux is low.  

The $v$-dependence of the cross sections in Figs. \ref{fig:muon-vdsdv}
and \ref{fig:stau-vdsdv}
may open a window for differentiating muons and staus in the
future. Fig. \ref{fig:stau-vdsdv} shows a marked difference between
muons and staus for large $v$ (small $1/v$). 
An energy deposit of
10\% of the initial stau energy of $E_\stau = 10^8$ GeV is observed
in a stau signal interaction, with very little in the way of comparable
energy losses nearby since the large $v$ cross section is so low.
With such a large energy loss (say, $\Delta E = 0.1E_\stau$), there would be
no  confusing the stau with muon with an energy of $E_\mu
= m_\mu/m_\stau\cdot E_\stau$ since $E_\mu\ll \Delta E$.
On the other hand, a similar energy deposit could come for a muon with the
same high energy, but the profile of the track would be significantly
different, with more large 
energy losses along the muon track where the multiple
pair production energy losses contribute. Essential here is the
large ratio of stau to muon masses. The effect is more difficult
to observe when
the masses are closer, e.g., for taus compared to muons \cite{montaruli}.

\begin{figure}[t]
\begin{center}
\epsfig{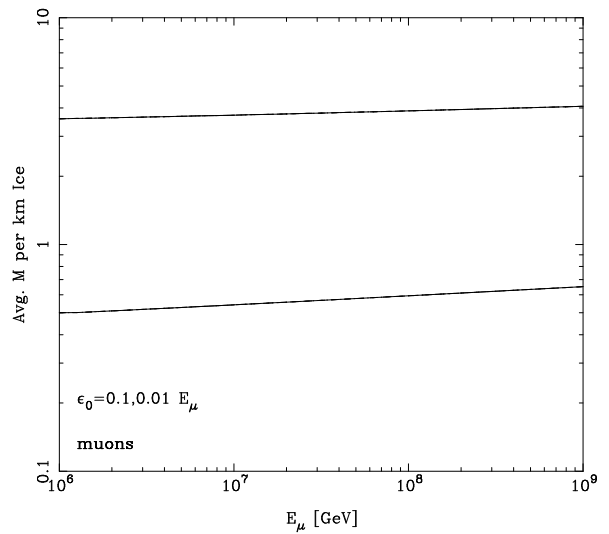}
\end{center}
\caption{The average number of interactions per km of ice for
muons. The upper line is for $v>0.01$,
and the lower line is for $v>0.1$.
}
\label{fig:fig3muon}
\end{figure}

\begin{figure}[t]
\begin{center}
\epsfig{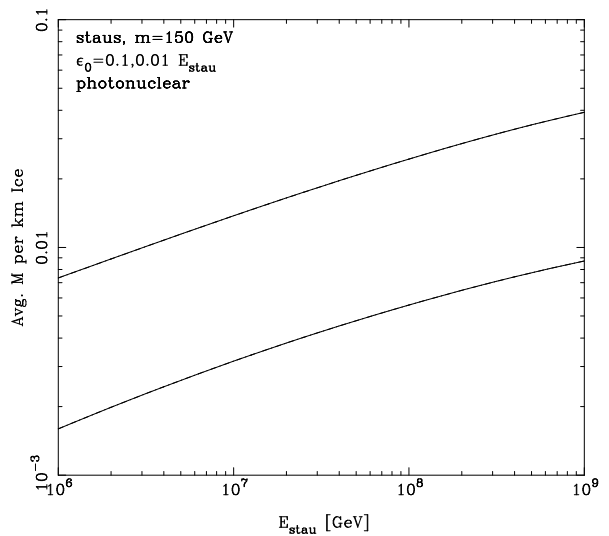}
\end{center}
\caption{The average number of interactions per km of ice for
staus with $m_\stau=150$ GeV. The upper line is for $v>0.01$,
and the lower line is for $v>0.1$.
}
\label{fig:fig3stau}
\end{figure}

In Figs. \ref{fig:fig3muon} and \ref{fig:fig3stau} we show the
average number of interactions per unit distance of ice rather than
the energy loss per unit distance. This is evaluated using
\begin{equation}
M = \frac{X\rho N_A}{A}\int_{v_{\rm min}}^1 dv\, \frac{d\sigma}{dv}
\ ,
\end{equation}
for density $\rho$ and column depth $X$.
Here we have taken
$v_{\rm min} = \epsilon_0/E=0.1, 0.01$ for muons and staus.
The lower line on  Figs. \ref{fig:fig3muon} and \ref{fig:fig3stau}
are for $v_{\rm min}=0.1$ and the upper lines are for
$v_{\rm min}=0.01$.
A muon which can deposit 10\% of its initial energy of $10^8$ GeV
proceeds to have a significantly larger number of subsequent
interactions in which 10\% of its energy is deposited, as
compared to the stau of the same energy. The scale of the average
number of interactions is such that one would need an
observational coverage of a few tens of
kilometer distance scales to see the distinction between
muons and staus. 
%{\it Some works to the effect that one needs to
%think about this issue for future expts.}

\subsection{Stau Showers in Ice}

Staus reaching the detector could interact 
in principle via weak interactions
producing showers. Weak interactions
of staus reduce its range \cite{RSU1}, but provide an
additional opportunity
for its detection via interactions in ice which can provide
a signal for detectors such as
ANITA \cite{anita} and
ARIANNA \cite{arianna}.  We discuss here showers produced by staus and the
background due to neutrino induced showers.  

We have evaluated stau and neutrino showers for different nadir
angles for energies above $10^6$ GeV.  In the shower production, we
assume maximal mixing for the 
charged-current interactions for staus.
The stau showers for
a given incident angle are
determined by Eq. (\ref{eq:ftaubc}), modified to include 
the probability to produce showers in the ice,
\begin{eqnarray}
\nonumber
F_{\stau,{\rm shr}}(E_{\rm shr},L)  &= &  
\int_{0}^{z_{ice}} F_\stau(E_{\rm shr},L-z') e^{
-{z'}/{{\cal L}_{CC}^{\stau}}}\frac{dz'}{{\cal L}_{CC}^\stau}
\\ \nonumber &\simeq&
F_\stau (E_{\rm shr},L)(1-e^{-z_{ice}/{\cal L}_{CC}^\stau})
\ .
\label{eq:showerstau}
\end{eqnarray}
The stau interaction length due to charged current interactions
is ${\cal L}_{CC}^\stau$. We use the
interaction length for maximal
weak interactions shown in Fig. \ref{fig:intlength}.
The pathlength through the ice, $z_{ice}$, for $\theta < 88.56^o$ 
is given by
\begin{eqnarray}
\label{eq:zice}
z_{ice} =  R_E\cos\theta - \sqrt{R_E^2\cos^2\theta - 2R_Et+t^2}, \quad
\end{eqnarray}
where $R_E$ represents the radius of the earth and $t$ the average
ice thickness, taken to be 2 km.  For comparison, showers due to neutrinos
are found from the equation,
\begin{equation}
\label{eq:showernu}
F_{\nu,{\rm shr}}(E_{\rm shr},L) 
\simeq  F_\nu(E_{\rm shr},L)(1-e^{-z_{ice}/{\cal L}_{CC}^\nu}) \ .
\end{equation}

\begin{figure}[t]
\begin{center}
\epsfig{file=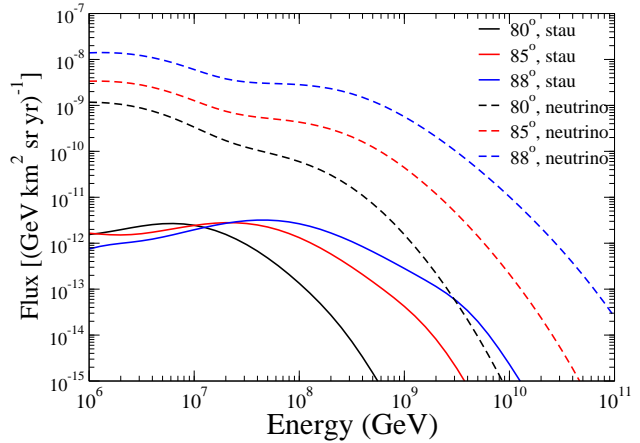,width=2.75in,angle=270}
\end{center}
\caption{The shower flux 
for the incident angles $80^\circ$ (lower curves), $85^\circ$ (middle curves), and 
$88^\circ$ (upper curves)  
from staus (solid lines) and neutrinos (dashed lines).}
\label{fig:showers}
\end{figure}

\begin{figure}[t]
\begin{center}
\epsfig{file=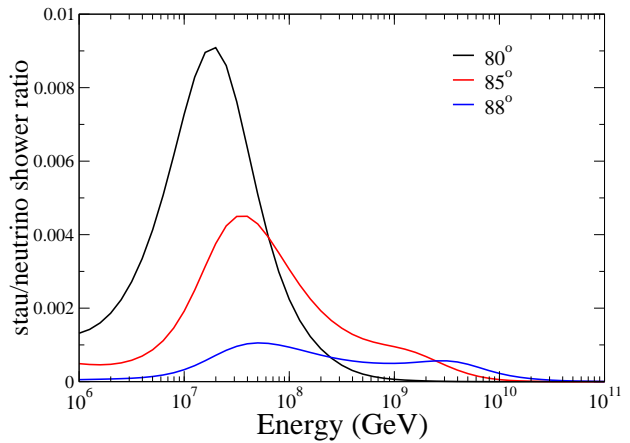,width=2.75in,angle=270}
\end{center}
\caption{The ratio of stau to neutrino induced shower fluxes 
for the incident angles $80^\circ$, $85^\circ$, and 
$88^\circ$, represented by curves with the highest, intermediate
and lowest peaks, respectively.}
\label{fig:shwrat}
\end{figure}

We compare the results for the showers due to staus and neutrinos for 
three nadir angles: $80^\circ$, 
$85^\circ$, and $88^\circ$ in Figs. \ref{fig:showers} and
\ref{fig:shwrat}.  All of the results in this section were obtained
using the standard evolution cosmogenic flux.

In Fig. \ref{fig:showers}, the showers due to neutrinos are of
the order $10^3-10^4$ larger than staus at an energy of $10^6$ GeV.
The input neutrino flux for the
stau signal will be at a higher energy than for the direct neutrino
production of showers.  The neutrino flux falls off as a
function of energy so the ratio of stau to neutrino showers will be
suppressed in part due to this effect. In addition, 
the stau flux is suppressed due to the small stau
production cross section.
The probability of producing showers in the detector is roughly
the same for neutrinos and for staus with maximal charged-current
interactions, so the inclusion of the showers is not sufficient to make up
for the suppression.

The
shape of the stau induced shower flux changes 
relative to the neutrino induced shower flux,
so between
$10^7-10^8$ GeV there is a peak in
the ratio of stau to neutrino induced showers.  
This effect is seen in Fig. \ref{fig:shwrat}, 
where the peak in the ratio of
stau/neutrino showers appears, but as indicated, the ratio is
small. It
is about
0.009 for $80^\circ$.  

\begin{figure}[t]
\begin{center}
\epsfig{file=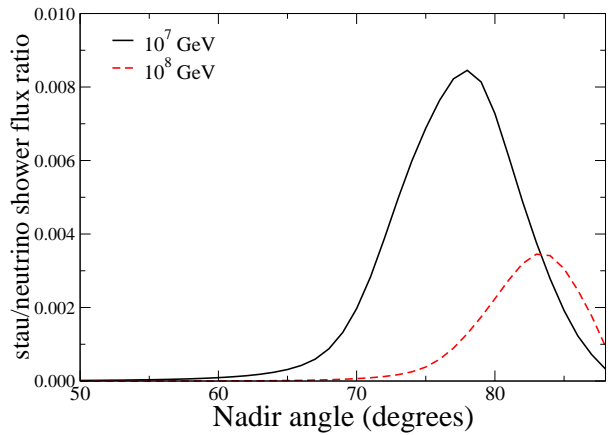,width=2.75in,angle=270}
\end{center}
\caption{The ratio of the fluxes of showers due to
stau and neutrino fluxes for the energies $10^7$ and $10^8$ GeV as a function of incident
angle.}
\label{fig:shwangrat}
\end{figure}

\begin{figure}[t]
\begin{center}
\epsfig{file=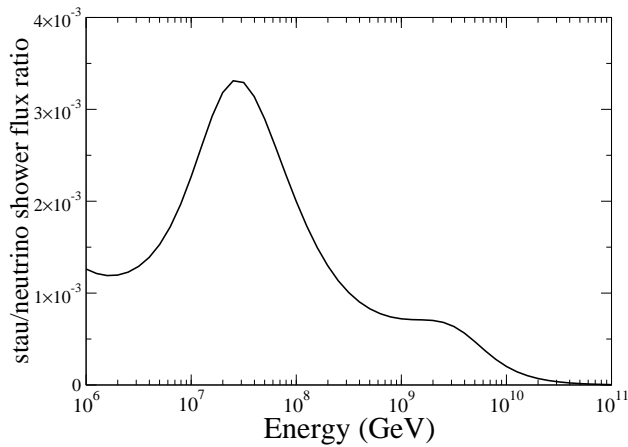,width=2.75in,angle=270}
\end{center}
\caption{The ratio of stau induced to neutrino induced shower
fluxes, integrated over all incident angles.}
\label{fig:totshower}
\end{figure}

Fig. \ref{fig:shwangrat} shows the stau/neutrino
ratio for the two fixed shower energies, $10^7$ and $10^8$ GeV,
 as a function of
nadir angle.  We can see that the ratio is the largest for $10^7$ GeV at
about $78^\circ$.  
The integrated
flux taking into account the contribution
due
to all incident angles is given in Fig. \ref{fig:totshower}.
The maximum stau/neutrino
ratio occurs for an energy of $2.5\times10^7$ GeV and corresponds to a stau
signal only 0.33 \% of the neutrino signal.

Whereas muon-like signals from staus can be large compared
to muon signals from neutrinos because of the long lifetime
of the stau,
we find that the shower signal
to background ratio is quite small for all angles.
This is due to combination 
of the
small production cross section for stau and the energy effects
due to the chain of production and the energy loss of the stau.
Stau attenuation
as it traverses the earth is roughly of the same size as
neutrino attenuation, as are the probabilities for staus or 
neutrinos 
to produce showers.

\section{Conclusions}

We have studied signals of staus produced in interactions of
cosmogenic neutrinos.  We have considered two types of signals, muon-like
charged tracks and showers.   We have focused on low scale supersymmetric models
that have stau as NLSP, which decays into the lightest SUSY particle, the gravitino.  
For a sufficiently large scale of supersymmetry breaking, the
stau has a very long lifetime.   

Our focus has been on cosmogenic neutrino fluxes and their associated
stau production in the Earth. Energy losses, both through electromagnetic
and weak interactions, are important in evaluating stau signals.
The energy loss of staus, however, is relatively
small in comparison with muons.  Thus, for some nadir angles, the stau flux is
much larger than the muon flux produced in neutrino charged-current interactions.

The enhancement of the stau flux is larger from an input
cosmogenic neutrino flux than for the Waxman-Bahcall
neutrino flux \cite{WB}. This is because the cosmogenic
neutrino flux is peaked at energies of about $10^8$ GeV, while the
WB flux is characterized by a steep power law with index of two.
The large ratio of staus to muons from cosmogenic neutrinos
is encouraging for experimental
detection, but in order to see this signal one needs to be able to
distinguish between staus and muons. Using the average energy loss
per unit distance is not a good way to distinguish staus and muons,
since the scaling of the energy loss parameter $\beta$
has the effect of making a high-energy stau look like a
lower energy muon.  We have proposed 
a way to distinguish between stau
and muon tracks by
measuring the energy loss of muons via
their interactions in the ice, and to use this method to reduce the background.

We also considered showers produced by staus interacting in the ice via charged-current
interactions.  The backgrounds for 
this signal are showers induced directly by
neutrinos that reach the detector and interact inside the detector via charged-current
or neutral-current interactions.  The only way that staus would produce showers in the
ice is if there is a weak mixing, but this process also contributes to reducing stau
range.  These effects combine with the small stau production probability to 
give fluxes of attenuated staus that are several orders of magnitude less than
attenuated neutrino fluxes. 
This small stau to neutrino ratio translates directly to shower
rates. 

In addition to weak interactions,
another possibility for shower production 
would be stau decays in the detector. For the parameter space considered
here, with long-lived staus, the signal from
decays is suppressed relative to their
weak interactions.  

In conclusion, stau signals at high energies are best identified by
muon-like tracks. The most important detection issue is distinguishing
between staus and muons, which may be possible by looking at the
incremental 
electromagnetic energy loss as the charged particle moves through
the detection volume. With very large volumes,
there is
a potential for detection of staus with future large neutrino
telescopes.

\acknowledgments

This work was supported in part by DOE contracts DE-FG02-91ER40664,
DE-FG02-04ER41319 and DE-FG02-04ER41298 (Task C). We thank A. Bulmahn
for discussions.

\end{document}